\documentclass[aps,pra,twocolumn,groupedaddress,notitlepage]{revtex4-1}
\usepackage{CJK}
\usepackage[pdftex]{graphicx}
\usepackage{float}
\usepackage{amssymb}
\usepackage{amsmath}
\usepackage{url}
\usepackage{natbib}
 
\usepackage[colorlinks,
                   linkcolor=blue,
                   anchorcolor=blue,
                   citecolor=blue,
                   urlcolor=blue]
                  {hyperref}              
\begin{document}
\title{Tri-spin dynamics in alkali metal-noble gas NMR gyroscope}
\author{Guobin Liu}
\email[Email:\,] {gbliu17@gmail.com}
\affiliation{Department of Physics, Ben-Gurion University of the Negev, Beer-Sheva 84105, Israel.}
\date{\today}
\begin{abstract}
Alkali metal-noble gas NMR gyroscope is widely used for precision rotation measurement in fundamental and applied physics. By numerically simulating the alkali-nuclear-nuclear tri-spin dynamics, we investigate the dependence of gyroscope response on alkali spin relaxation time and nuclear spin magnetization. We found additional resonance peaks appear due to a new source of instability, namely the inherent multistability of tri-spin dynamics. The numerical simulation results agree well with the recent experiment, enabling a better understanding and exploitation of the gyroscope signal.
\end{abstract}
\pacs{}
\maketitle

\section{introduction}
A class of precision measurement in physics relies on measuring the rotational inertial effect of spin precession, i.e., the measured Larmor frequency is shifted to $\omega\,'$=$\omega$$\pm$$\Omega_R$, with $\omega$ the intrinsic Larmor frequency and $\Omega_{\rm R}$ the inertial rotation rate of the experimental setup carrying the spin. NMR signal of long lived nuclear spin polarization is utilized to extract $\omega\,'$. Such NMR gyroscope or comagnetometry has been widely used for experimental test of Lorentz invariance \cite{Larmoreaux1986PRL,Gemmel2010PRD,Smiciklas2011PRL}, search for CP-violating electric dipole moment \cite{Rosenberry2001PRL}, anomalous spin couplings beyond the standard model \cite{Kornack2005PRL}, precision rotation sensor \cite{Donley2010NMRG} and as unit sensor in the global network searching for axion \cite{Kimball2018PRD}.

One kind of NMR gyroscope uses alkali metal-noble gas dual-spin ensemble in a thermal vapor cell, where alkali spin serves both as a medium for polarization and as an embedded probe of the noble gas nuclear spin (over all electron spin vanishes due to full shell in electronic configuration). The gyroscope signal can be greatly enhanced due to the Fermi contact between alkali metal electronic and noble gas nuclear spins \cite{Schaefer1989PRA,Kornack2005PRL}. 

The simplest alkali metal-noble gas ($A$-$N$) gyroscope is based on dual spin comagnetometry \cite{Kornack2005PRL}, which works in a well compensated zero magnetic field. In this case, the gyroscope signal is insensitive to any ambient magnetic field and only sensitive to the earth rotation or anomalous interactions coupling to nuclear or electron spins. However, periodic calibration for the compensation magnetic field is needed to maintain a zero compensation point \cite{Smiciklas2011PRL}, making a continuous measurement impossible.

Spatial and temporal variations of the external field ${\bf B}_0$ in which the spins precess limit gyroscope sensitivity. By introducing an additional noble gas, one can measure simultaneously the precession signal of two nuclear spins occupying the same volume \cite{Donley2010NMRG,Korver2015PRL,Limes2018PRL}, thus $\Omega_R$ can be determined without knowing ${\bf B}_0$ via 
\begin{equation}
\Omega_{\rm R}=\frac{\gamma_1\omega_2\,'-\gamma_2\omega_1\,'}{\gamma_1-\gamma_2},
\label{rotation}
\end{equation} 
with $\gamma_{1, 2}$ the gyromagnetic ratio of nuclear spin $N_1$ and $N_2$, respectively. This alkali metal-noble gas-noble gas ($A$-$N_1$-$N_2$) tri-spin NMR gyroscope can work in a finite nonzero magnetic field and thus loose the requirement on maintaining a well controlled zero ambient magnetic field, making a long term continuous measurement possible \cite{Limes2018PRL}. The long term continuous measurement is important for practical applications where the calibration of compensation field is unavailable or increasing the complexity of the apparatus if difficult. An additional benefit of the tri-spin gyroscope is its measurement independence on the orientation of the gyroscope \cite{Limes2018PRL}, compared to the dual spin version \cite{Kornack2005PRL}.

An alkali spin magnetometry usually measures dc or slowly varying magnetic field. Steep Zeeman resonance is utilized to monitor the change of field within the resonance linewidth. In tri-spin NMR gyroscope, however, the signal field comes from two nuclear magnetizations in motion and alkali spin probes it via coupling to nuclear spins. Besides, from classical point of view, alkali spin relaxation time determines the Zeeman resonance linewidth $\Delta\nu$ and thus the dynamic range of alkali spin magnetometry $\Delta B$ via 
\begin{equation}
\begin{split}
\Delta B=\frac{\Delta\nu}{\gamma_{\rm A}}=\frac{1}{\pi \gamma_{\rm A} T^{\rm A}}.
\end{split}
\label{dr}
\end{equation}
With typical alkali gyromagnetic ratio $\gamma_{\rm A}$$\sim$1\,MHz/G and spin relaxation time $T^{\rm A}$$\sim$1\,ms, the alkali spin magnetometry has a finite dynamic range $\Delta B$$\sim$0.3 mG. The signal of nuclear magnetization is typically at the same level or higher \cite{Limes2018PRL,Kornack2002PRL}. It is thus necessary to study the response of gyroscope and its dependence on alkali spin relaxation time and nuclear magnetization. 

We have noticed that recently Limes $et$ $al$. observed a multiple resonance spectrum, i.e., the so called ``cross-modulation'' peaks in a Rb-He-Xe tri-spin system \cite{Limes2018PRL} and they concluded that a better understanding is necessary in order to reach the full sensitivity potential. Here we try to understand the experimental phenomenon and related effects by a numerical simulation of the tri-spin dynamic system.

\section{Coupled Bloch equations and numerical solutions} 

We simulate the spin dynamics of the $A$-$N_1$-$N_2$ tri-spin system via coupled Bloch equations as was done for the $A$-$N$ dual-spin system by Kornack $et$ $al$. \cite{Kornack2002PRL}. 

First we assume that there is a classical macroscopic magnetization field $\bf M$ for both alkali vapor and noble gases after the spin-exchange optical pumping. Two fundamental magnetic interactions dominate the spin magnetization dynamics of the tri-spin system, the Zeeman interaction between atomic spins and the bias field ${\bf B}_0$ and spin-exchange collisional interaction between alkali vapor and noble gas spins. The first interaction is well described by NMR Bloch equation. For the second one, Schaefer $et$ $al$. have shown that EPR (electron paramagnetic resonance) and NMR frequency shift happen to alkali and noble gas spin resonance, respectively, due to the short lived but enhanced spin-exchange interactions \cite{Schaefer1989PRA}. These frequency shifts can be simulated by introducing an additional effective magnetic field, a spin-exchange field $\bf B_{se}$. Due to the attractive force between alkali electron and noble gas nucleus, $\bf B_{se}$ is enhanced compared to the classical macroscopic magnetization field by a factor $\lambda$, thus we have
\begin{equation}
{\bf B_{se}}=\lambda {\bf M}=\frac{8 \pi}{3}\kappa_0 \bf M,
\label{bse}
\end{equation}
with $\kappa_0$ the enhancement factor over the classical magnetic field due to the attraction of alkali electron wave function to noble gas nucleus \cite{Schaefer1989PRA,Kornack2002PRL}. 

Based on above assumption, we can thus add these spin-exchange fields into the classical Bloch equation and form a set of coupled Bloch equations. We consider first the basic configuration of tri-spin dynamics where all magnetic fields and magnetizations couple to each other. We refer it as complete-coupling configuration, for which the equations of motion can be written as
\begin{equation}
\small
\begin{split}
\frac{\partial{\bf M}^{\rm A}}{\partial t}&=\frac{\gamma_{\rm A}}{q}{{\bf M}^{\rm A}}\times[{\bf B}_0+\lambda_1 {{\bf M}^{\rm N_1}}+\lambda_2{{\bf M}^{\rm N_2}}] +\frac {M_0^{\rm A}{\hat z}-{\bf M}^{\rm A}} {\it{q T^{\rm A}}},\\
\frac{\partial {\bf M}^{\rm N_1}}{\partial t}&=\gamma_1{{\bf M}^{\rm N_1}}\times[{\bf B}_0+\lambda_1{\bf M}^{\rm A}+\lambda_3{\bf M}^{\rm N_2}]+\frac {{\it M}_0^{\rm N_1}{\hat z}-{\bf M}^{\rm N_1}} {[{\it T}_2^{\rm N_1}, {\it T}_2^{\rm N_1}, {\it T}_1^{\rm N_1}]},\\
\frac{\partial {\bf M}^{\rm N_2}}{\partial t}&=\gamma_2{{\bf M}^{\rm N_2}}\times[{\bf B}_0+\lambda_2{\bf M}^{\rm A}+\lambda_3 {{\bf M}^{\rm N_1}}]+\frac{M_0^{\rm N_2}{\hat z}-{{\bf M}^{\rm N_2}}} {[{\it T}_2^{\rm N_2}, {\it T}_2^{\rm N_2},{\it T}_1^{\rm N_2}]},
\label{cbe0}
\end{split}
\end{equation}
where $\gamma$ with subscripts are gyromagnetic ratios for corresponding spins. ${\bf M}^{\rm A}$ and ${\bf M}^{\rm N_1/N_2}$ are the alkali vapor and noble gas magnetizations, respectively. $\lambda_{\rm 1}, \lambda_{\rm 2}$ and $\lambda_{\rm 3}$ are enhancement coefficients for $A$-$N_1$, $A$-$N_2$ and $N_1$-$N_2$ spin couplings, respectively. We assume that alkali spin has the same timescale for both longitudinal and transverse relaxation times, termed as alkali spin relaxation time $T^{\rm A}$. $q$ in the first equation is the slowing down factor depending on the alkali spin polarization, ranging from 6 for low polarization to 4 for high polarization \cite{Savukov2005PRA}. In the square brackets of last two equations, ${\it T}_2^{\rm N_1/N_2}$ and ${\it T}_1^{\rm N_1/N_2}$ represent the transverse ($M_{x, y}$) and longitudinal ($M_z$) relaxation times, respectively. 

So far, analytical solution does not exist for the three-body dynamical system, except for some special cases \cite{3body2013PRL}. We thus resort to numerical solutions and use the embedded solver ODE45 of MATLAB software to solve the coupled Bloch equations. We assign specific species for the the $A$-$N_1$-$N_2$ spin system, for example, rubidium-87 for $A$, helium-3 for $N_1$ and xenon-129 for $N_2$. Typical values for gyroscope are used for simulation \cite{Kornack2002PRL, Limes2018PRL}. For example, $^3$He/$^{129}$Xe spin magnetization field felt by alkali spin is $\sim$20\,$\mu$G-1\,mG and $^{87}$Rb spin magnetization felt by nuclear spin is $\sim$20\,$\mu$G. We assume that $N_1$ and $N_2$ initial magnetization fields are equal, termed as $\lambda M_0^{\rm N}$. We have checked that the difference between $N_1$ and $N_2$ initial magnetization only changes the relative strength of the two resonance signal but not the general results as will be described later. The dc bias field ${\bf B}_0$ is along the $z$ axis. The initial He and Xe spin magnetizations are set along and against the $y$ axis, respectively. The initial alkali spin magnetization is set along the $z$ axis. The $T_1$ and $T_2$ relaxation times are taken 1\,hour and 1000\,s for He spin magnetization, 1000\,s and 80\,s for Xe spin magnetization. The enhancement factor $\kappa_0$ is taken as 5, 500 and -0.011 for Rb-He \cite{Romalis1998PRA}, Rb-Xe \cite{Ma2011PRL} and He-Xe \cite{Limes2018arxiv} spin exchange interaction, respectively.

\begin{figure}[H]
\centering
\includegraphics[width=\columnwidth]{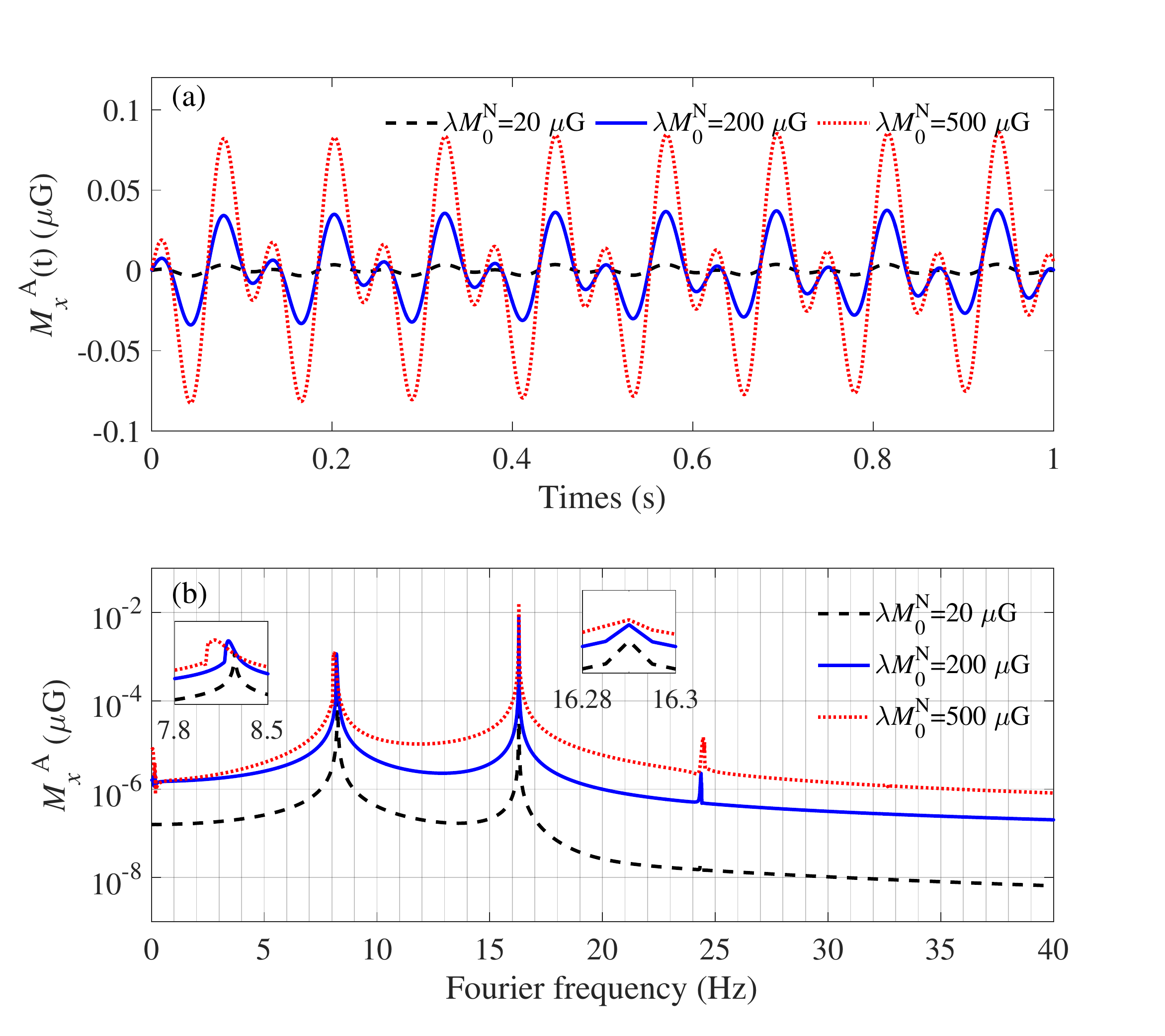}
\caption{Gyroscope signal $M_x^{\rm A}$ versus time (a) and frequency (b) for different initial nuclear magnetizations simulated for the complete coupling tri-spin system, with alkali spin relaxation time $T^{\rm A}$=1\,ms, meaning a dynamic range $\Delta B$=0.45\,mG given $\gamma^{\rm A}$=0.7\,MHz/G for $^{87}$Rb, based on Eq.\,\ref{dr}.}
\label{fig1}
\end{figure} 

The sensing medium of NMR gyroscope is alkali spin polarization, whose transverse component is read out by optical method as the gyroscope signal \cite{Kornack2002PRL,Kornack2005PRL,Limes2018PRL}. Fig.\,\ref{fig1}(a) shows gyroscope signal $M_x^{\rm A}$(t) over one second, for different initial nuclear magnetizations $\lambda M_0^{\rm N}$. Due to a much shorter relaxation time, the alkali spin can adiabatically follow up the precessing nuclear magnetization. The alkali spin follows the slow precession of two nuclear spins around a dc bias field, resembling a dual-frequency oscillation. The gyroscope signal increases with the nuclear magnetization. A slow drift appears in the long term evolution in Fig.\,\ref{fig1}(a), which corresponds to a dc component in frequency spectrum shown in Fig\,\ref{fig1}(b). The spectrum is Fourier transformation of oscillation over 200 seconds. 

The main resonance peaks at around 8.21\,Hz and 16.29\,Hz are the $^{129}$Xe and $^3$He spin resonances. Due to spin exchange interaction \cite{Schaefer1989PRA}, the nuclear magnetization cause a frequency shift to alkali spin precession and vice versa. The Xe-peak is obviously shifted (relative to intrinsic frequency of 5.895\,Hz at $B_0$=$|{\bf B}_0|$=5\,mG) due to the strong Rb-Xe spin exchange interaction with $\kappa_0^{\rm Rb-Xe}$$\sim$500, while the He-peak is only slightly shifted (relative to 16.225\,Hz at $B_0$=5\,mG) due to $\kappa_0^{\rm Rb-He}$$\sim$5. The $N_1$-$N_2$ spin coupling is negligible due to much smaller $\kappa_0^{\rm Xe-He}$. The positions of main resonance peaks depend on nuclear magnetization (insets of Fig.\,\ref{fig1}(b)) and relaxation time (see Fig.\,\ref{fig7}(b) in the Appendix). In conclusion, the complete coupling tri-spin system is unsuitable for precision NMR gyroscope. 

We consider an ideal configuration of tri-spin dynamics where: 1) Nuclear spins precess solely due to the bias field along the $z$ axis; 2) Alkali spin precesses solely due to the nuclear magnetizations. Technically, the NMR frequency shift of nuclear spins caused by alkali spin can be averaged out by the so called dynamical decoupling method \cite{Korver2015PRL, Limes2018PRL}. The precession of alkali spins around the dc bias field can also be averaged out by fast periodic modulation on the optical pumping \cite{Limes2018PRL}. For this partial coupling configuration, a new set of coupled Bloch equations can be written as
\begin{equation}
\small
\begin{split}
\frac{\partial {\bf M}^{\rm A}}{\partial t}&=\frac{\gamma_{\rm A}}{q}{{\bf M}^{\rm A}}\times[\lambda_1 {{\bf M}^{\rm N_1}}+\lambda_2{{\bf M}^{\rm N_2}}] +\frac {M_0^{\rm A}{\hat z}-{\bf M}^{\rm A}} {\it{q T^{\rm A}}},\\
\frac{\partial {\bf M}^{\rm N_1}}{\partial t}&=\gamma_1{{\bf M}^{\rm N_1}}\times[{\bf B}_0+\lambda_3{\bf M}^{\rm N_2}]+\frac {{\it M}_0^{\rm N_1}{\hat z}-{\bf M}^{\rm N_1}} {[{\it T}_2^{\rm N_1}, {\it T}_2^{\rm N_1}, {\it T}_1^{\rm N_1}]},\\
\frac{\partial {\bf M}^{\rm N_2}}{\partial t}&=\gamma_2{{\bf M}^{\rm N_2}}\times[{\bf B}_0+\lambda_3 {{\bf M}^{\rm N_1}}]+\frac{M_0^{\rm N_2}{\hat z}-{{\bf M}^{\rm N_2}}} {[{\it T}_2^{\rm N_2}, {\it T}_2^{\rm N_2},{\it T}_1^{\rm N_2}]},
\label{cbe1}
\end{split}
\end{equation}

\begin{figure}[H]
\centering
\includegraphics[width=\columnwidth]{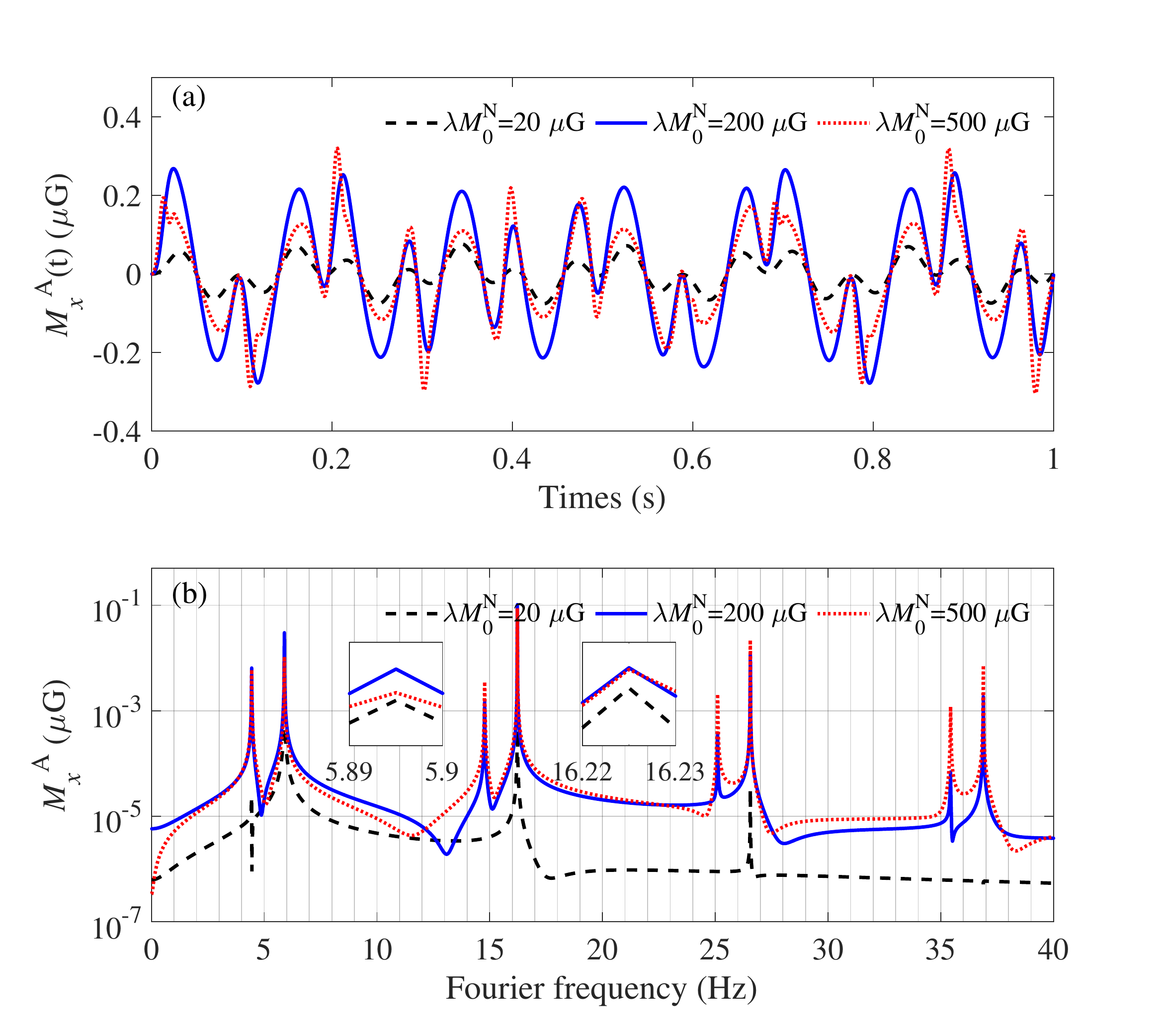}
\caption{Gyroscope signal versus time (a) and frequency (b) for different initial nuclear magnetizations simulated for the partial coupling tri-spin system, with alkali spin relaxation time $T^{\rm A}$=1\,ms.}
\label{fig2}
\end{figure}

Fig.\,\ref{fig2} shows the gyroscope signal simulated for the partial coupling tri-spin system with the same parameters as for the complete coupling configuration. In temporal signal, although the phase of dual-frequency oscillation distorts as the nuclear magnetization increases, the overall oscillation is dominated by two frequencies. This is proved by Fourier analysis of oscillations over 200 seconds, as shown in Fig.\,\ref{fig2}(b), two main resonance peaks independent of nuclear magnetization exist at 5.895\,Hz and 16.225\,Hz (see inset). We also show that they are independent of alkali spin relaxation time (see Fig.\,\ref{fig8} in the Appendix). In conclusion, the partial coupling tri-spin system can ensure a precision NMR gyroscope based on Eq.\,\ref{rotation}.

Fourier analysis also shows that additional resonance peaks show up in the spectrum. A minor `associate' peak near the main Xe-peak appear first when the nuclear magnetization is small, i.e., when $\lambda M_0^{\rm N}$=20\,$\mu$G$\ll$$\Delta B$=0.45\,mG for the dashed black line in Fig.\,\ref{fig2}(b). As the nuclear magnetization increases, more peaks appear and grow with increasing $\lambda M_0^{\rm N}$. This could be harmful for gyroscope. For example, as the nuclear magnetization approaches the dynamic range of alkali magnetometry, the associated peak next to main resonance becomes comparable in amplitude to each other. In this case, it could be practically problematic to figure out the true value of $\omega\,'$ as the peaks are closely spaced in spectrum. 

We have noticed that it is difficult to measure accurately the nuclear spin polarization without a complicated apparatus \cite{Romalis1998PRA, Ma2011PRL}. Alternatively, as we have shown here, fitting simulated gyroscope signal to experimental data could be a good method to estimate quantitatively the nuclear magnetization strength or alkali spin relaxation time in situ.  

\section{origin of additional resonance peaks}

From our simulations we noticed that additional resonance peaks grow in quantity and amplitude as the nuclear magnetization or alkali spin relaxation time increases. However, it is still unclear why these additional peaks are created from the very beginning. 
We attribute it to the instability of three-body dynamics rooted in the tri-spin ensemble. To prove this, we look at the dynamical trajectories of alkali spin precession and compare them between the complete and the partial coupling configurations. 

\begin{figure}[H]
\centering
\includegraphics[width=\columnwidth]{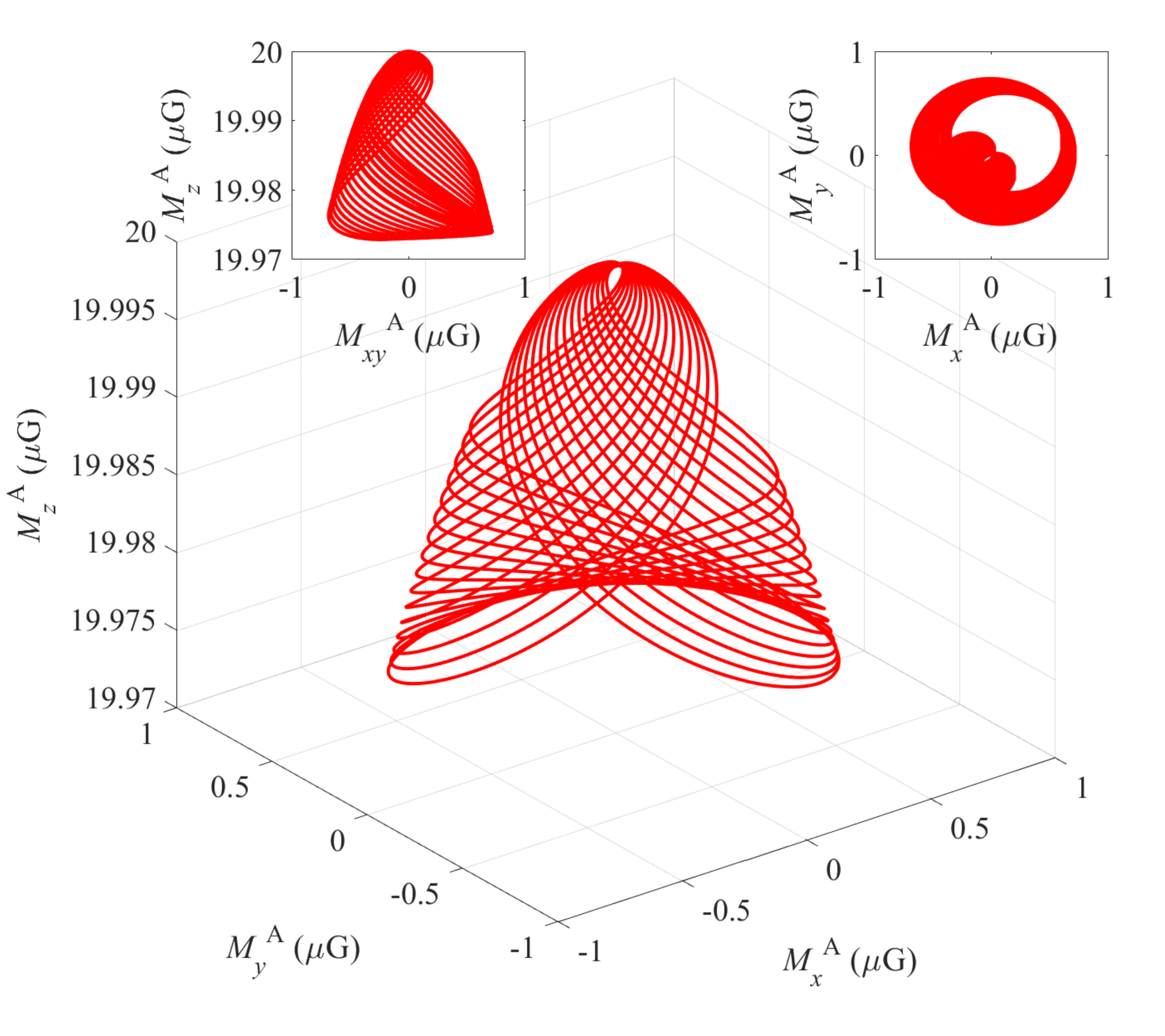}
\caption{Trajectory of alkali spin precession simulated for the complete coupling tri-spin system, with nuclear magnetization $\lambda M_0^{\rm N}$=0.1\,mG and alkali spin relaxation time $T^{\rm A}$=1\,ms.}
\label{fig3}
\end{figure}

For the complete coupling configuration,  Fig.\,\ref{fig3} shows the trajectory of alkali spin precession in the initial two seconds. Although both bias field and nuclear magnetization apply torques to the alkali spin, the former is stronger than the latter. Thus the alkali spin is only slightly tilted away from the quantum axis along the $z$ axis.
The fast precession of alkali spin around the quantum axis field is dominant. The torque applied by nuclear magnetization is small, acting on the alkali spin as a perturbation or weak modulation. As the two nuclear spins have different gyromagnetic ratios, the modulation is dual-frequency. Thus a double-ring structure forms in the trajectory, as shown in Fig.\,\ref{fig3}. 

The reasons that we show only the initial two seconds of trajectory are the following: 1) The nuclear spin precesses a few cycles within one second, so two seconds are enough to show the trend of dynamics in general; 2) At longer times, alkali spin trajectory will fill up the whole space due to its everlasting changing phase in tri-spin dynamics. This will smear the main structure of trajectory, hindering us from understanding the essences of different tri-spin dynamics.

For the range of nuclear magnetization studied here, i.e., $\lambda M_0^{\rm N}$ from 20\,$\mu$G to 0.5\,mG, the tilting angle of alkali spin is small and linearly proportional to $\lambda M_0^{\rm N}$. In this regime, the structure of trajectory does not change as nuclear magnetization increases (see Fig.\,\ref{fig9} in the Appendix). The gyroscope signal measuring the transverse alkali magnetization, is small and increases linearly with nuclear magnetization, consistent with the simulation results shown in Fig.\,\ref{fig1}. 

In contrast, for the partial coupling tri-spin dynamics, the trajectory is very different. Here the bias field no longer applies any torque to the alkali spin. The nuclear magnetization applies a torque dominating the motion of alkali spin, which at first is tilted at a large angle away from the quantum axis after which is growing into a nonlinear region. Then the alkali-nuclear tri-spin coupling dominates the spin dynamics. 

\begin{figure}[H]
\centering
\includegraphics[width=\columnwidth]{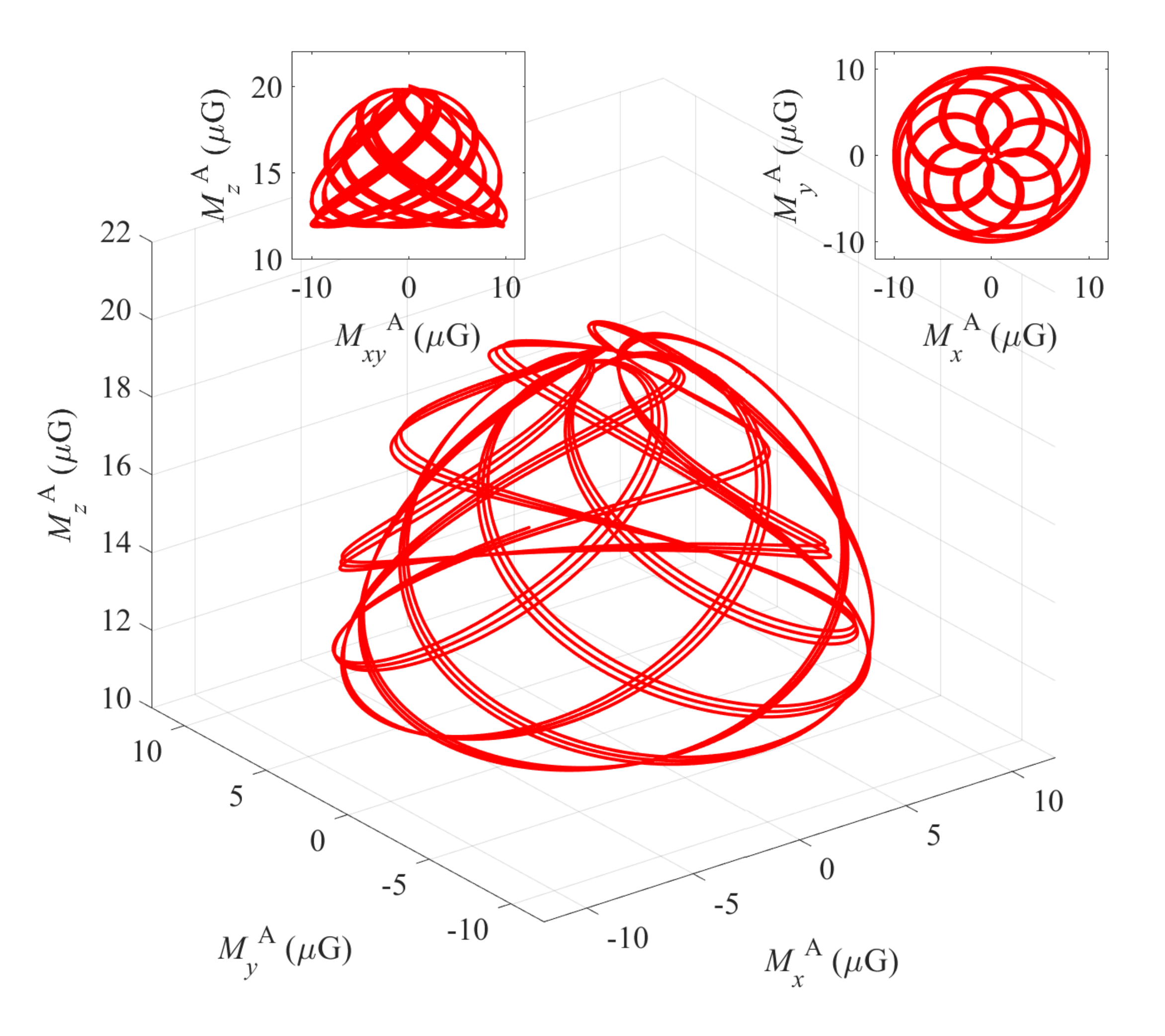}
\caption{Trajectory of alkali spin precession simulated for the partial coupling tri-spin system, with nuclear magnetization $\lambda M_0^{\rm N}$=0.1\,mG and alkali spin relaxation time $T^{\rm A}$=1\,ms.}
\label{fig4}
\end{figure}

It's well known that three-body dynamics has a complex behavior, resulting in unpredictable long term dynamics or even chaos (dynamically instable, accompanying with bistability or multistability) in some cases \cite{chaos2012}. The dominance of tri-spin dynamics leads to a similar complex trajectory of alkali spin precession, as shown in Fig.\,\ref{fig4}. First, more rings are present (up right inset) and secondly, as the nuclear magnetization increases, the structure of trajectory changes dramatically (see Fig.\,\ref{fig10} in the Appendix). One can notice that the complexity of trajectory is positively correlated to the number of additional resonance peaks in spectrum. 

\begin{figure}[H]
\centering
\includegraphics[width=\columnwidth]{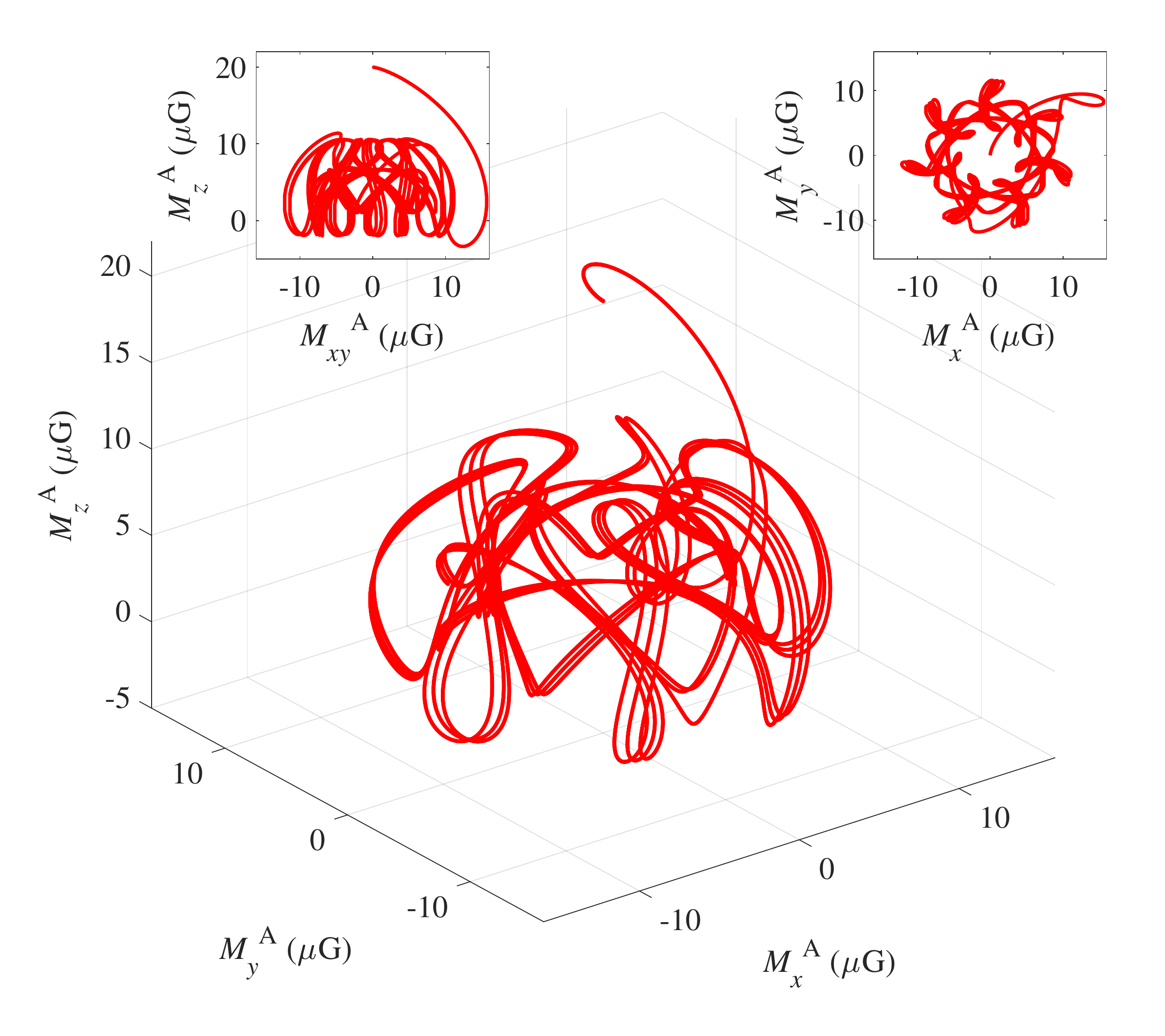}
\caption{Trajectory of alkali spin precession simulated for the partial coupling tri-spin system, with nuclear magnetization $\lambda M_0^{\rm N}$=0.1\,mG and alkali spin relaxation time $T^{\rm A}$=10\,ms.}
\label{fig5}
\end{figure}

The alkali spin relaxation time plays an even more significant role in the complexity of tri-spin dynamics. Fig.\,\ref{fig5} shows the structure of trajectory changing dramatically as $T^{\rm A}$ increase by an order of magnitude. For the same nuclear magnetization as in Fig.\,\ref{fig4}, but with ten times longer relaxation time, the alkali spin can flip many times across the $xy$ plane (the equatorial plane $z$=0, up left inset). The multiple crossovers with the equatorial plane is correlated again to the creation of more additional resonance peaks, as shown in Fig.\,\ref{fig8} (see the Appendix). 

Based on the above analysis, we will expect that the quantity and amplitude of additional resonance peaks grow with increasing tilting angle of alkali spin 
\begin{equation}
\theta=\gamma^{\rm A}\lambda M_{xy}^{\rm N}{\rm (t)}T^{\rm A}/q.
\label{theta}
\end{equation}
However, in principle these additional resonance peaks are inherent to the tri-spin system. In fact, they appear even for very small tilting angle, where the nuclear magnetization is far below the dynamic range determined by the alkali spin relaxation time. For example, the aforementioned `associate' peak shown in Fig.\,\ref{fig2}(b), where $\theta$ is 2\,$^\circ$ at most (when $N_1$ and $N_2$ spin magnetizations are along the same direction) with $\lambda M^{\rm N}_0$=20\,$\mu$G and $T^{\rm A}$=1\,ms.

\begin{figure}
\centering
\includegraphics[width=\columnwidth]{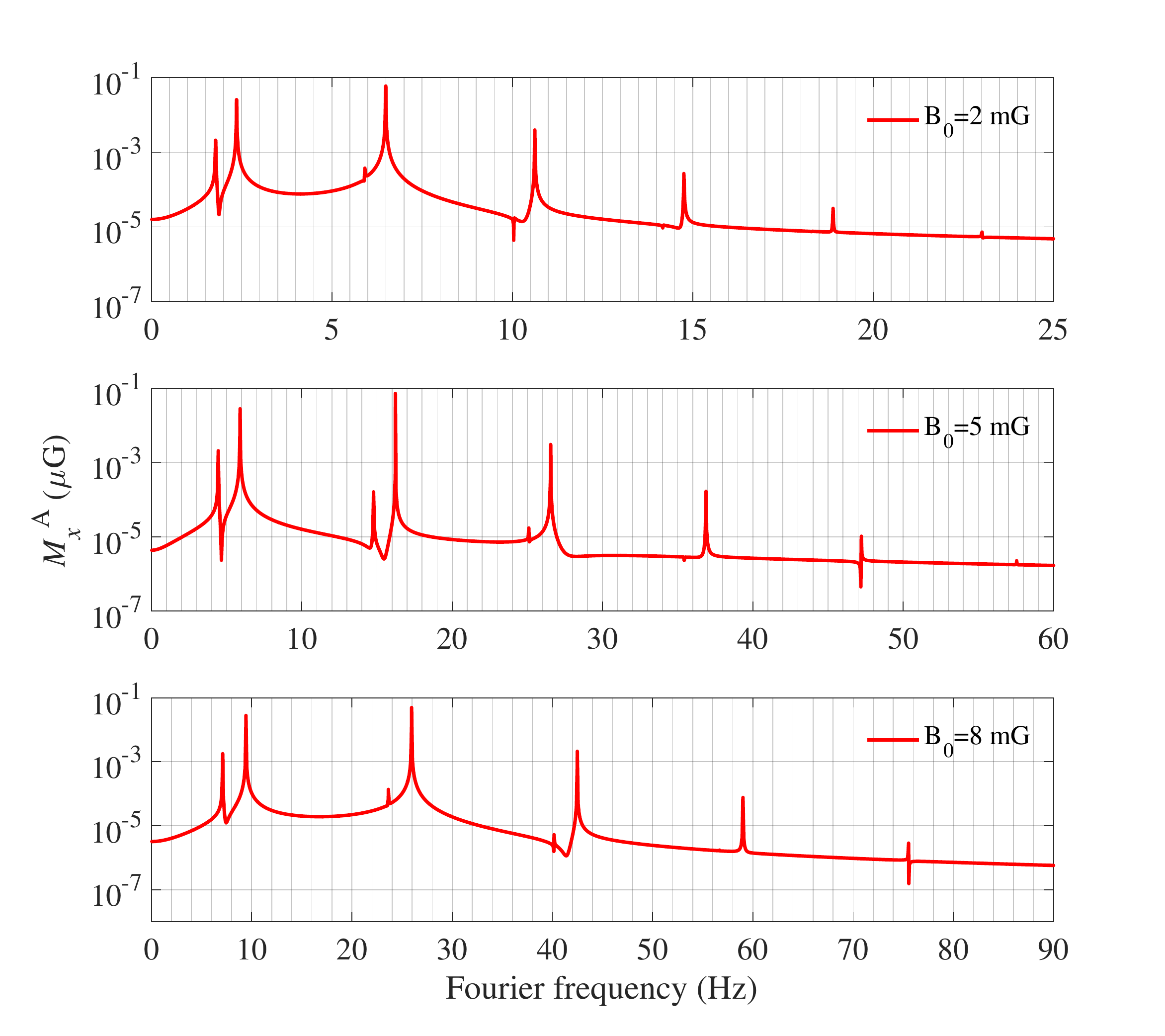}
\caption{Dependence of gyroscope spectral response on bias field simulated for the partial coupling tri-spin system, with nuclear magnetization $\lambda M_0^{\rm N}$=0.1\,mG and alkali spin relaxation time $T^{\rm A}$=1\,ms.}
\label{fig6}
\end{figure}

While the tri-spin dynamics is inherently complex to be expressed in any analytic form, we try to find out some empirical rules from numerical simulations. We noticed that there are two kinds of additional resonance peaks in spectrum. One kind appears at higher frequencies. The other kind appears as an associate peak closely spaced to the main peaks and to the first kind of additional peaks. We investigate the dependence of multiple resonance peaks on the bias field $B_0$=$|{\bf B}_0|$. Fig.\,\ref{fig6} shows the simulated spectrum versus $B_0$ from Eq.\,\ref{cbe1}. From the measured positions of multiple peaks, we found the following formulas
\begin{equation}
\begin{split}
\Delta&=(\gamma_1-\gamma_2)B_0,\\
\delta&=(\lceil {\gamma_1}/{\gamma_2} \rceil \gamma_2-\gamma_1) B_0,
\label{fs}
\end{split}
\end{equation}
where $\Delta$ is the differential frequency between two main resonance peaks of nuclear spin and $\delta$ is the distance between the main peaks and their nearest associate peaks. $\lceil {\gamma_1}/{\gamma_2} \rceil$ is rounding up to the next integer of the ratio between nuclear gyromagnetic ratios (assuming $\gamma_1$$>$$\gamma_2$). One may notice that once the gyromagnetic ratios of two nuclear species are determined, the ratio $\Delta$/$\delta$ is also settled. In the case of Rb-$^3$He-$^{129}$Xe system, $\Delta$/$\delta$$\approx$7.1241, meaning about six additional resonance peaks if equally spaced in between Xe and He main peaks. 

\section{restricted planar motion and multiple resonance peaks}

Based on the partial coupling configuration, we further consider a particular case where the alkali spin precession is restricted within the a plane due to some practical reasons. For example, the restricted planar motion of spin precession could be effectively realized, by a serie of experimental techniques such the synchronous optical pumping, NMR spin locking, or a combination of them or their slightly modificated versions. 

Actually, we have noticed that the modulated optical pumping and the pulse train technique in the experiment may effectively realize such a restricted planar motion for alkali spin polarization and lead to its sensitivity mainly to the nuclear magnetization $M_y^{\rm N}$ \cite{Limes2018PRL}. Therefore, any torque applied to alkali spin by nuclear magnetization lying in the $x$ or $z$ axis will be averaged out effectively, so the first equation of Eqs.\,\ref{cbe1} can be simplified by eliminating terms containing $M_x^{\rm N}$ or $M_z^{\rm N}$. The new set of equations of motion now becomes 
\begin{equation}
\small
\begin{split}
\frac{\partial{M}_x^{\rm A}}{\partial t}&=-\frac{\gamma_{\rm A}}{q} M_z^{\rm A}(\lambda_1 M_y^{\rm N_1}+\lambda_2 M_y^{\rm N_2}) -\frac {M_x^{\rm A}} {\it{q T^{\rm A}}},\\
\frac{\partial{M}_y^{\rm A}}{\partial t}&=-\frac {M_y^{\rm A}} {\it{q T^{\rm A}}},\\
\frac{\partial{M}_z^{\rm A}}{\partial t}&=\frac{\gamma_{\rm A}}{q} M_x^{\rm A}(\lambda_1 M_y^{\rm N_1}+\lambda_2 M_y^{\rm N_2}) +\frac {M_0^{\rm A}-M_z^{\rm A}} {\it{q T^{\rm A}}},\\
\frac{\partial {\bf M}^{\rm N_1}}{\partial t}&=\gamma_1{{\bf M}^{\rm N_1}}\times[{\bf B}_0+\lambda_3{\bf M}^{\rm N_2}]+\frac {{\it M}_0^{\rm N_1}{\hat z}-{\bf M}^{\rm N_1}} {[{\it T}_2^{\rm N_1}, {\it T}_2^{\rm N_1}, {\it T}_1^{\rm N_1}]},\\
\frac{\partial {\bf M}^{\rm N_2}}{\partial t}&=\gamma_2{{\bf M}^{\rm N_2}}\times[{\bf B}_0+\lambda_3 {{\bf M}^{\rm N_1}}]+\frac{M_0^{\rm N_2}{\hat z}-{{\bf M}^{\rm N_2}}} {[{\it T}_2^{\rm N_2}, {\it T}_2^{\rm N_2},{\it T}_1^{\rm N_2}]}.
\label{cbe2}
\end{split}
\end{equation}

\begin{figure}[H]
\centering
\includegraphics[width=\columnwidth]{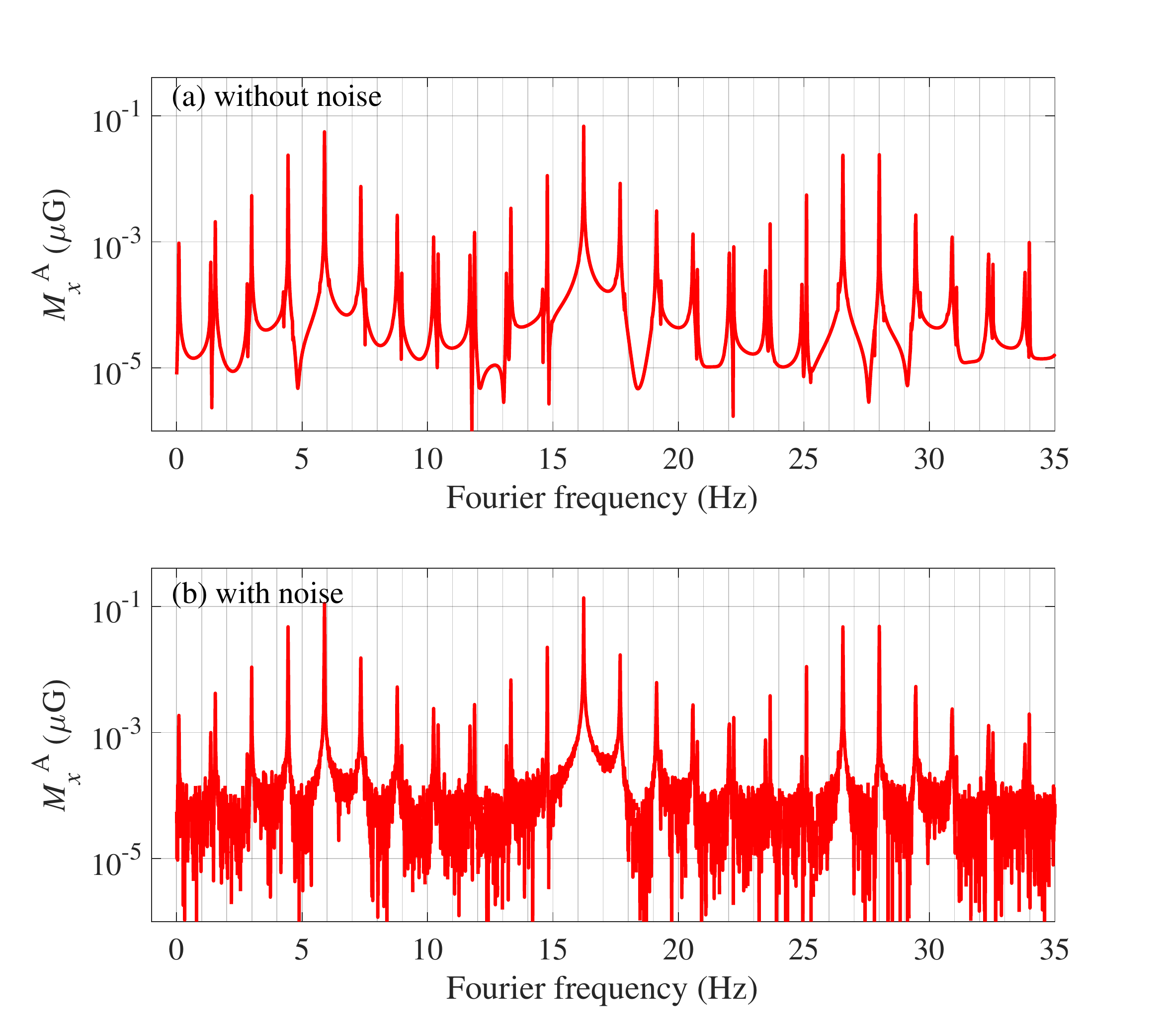}
\caption{Gyroscope spectral response simulated for planar-restricted-motion of the Rb-$^3$He-$^{129}$Xe tri-spin system. (a) and (b) are results without and with white Gaussian noise (signal to noise ratio is 10$^6$), respectively. The bias field is $B_0$=5\,mG. The He and Xe magnetization fields are 0.5\,mG and 1.5\,mG, respectively. The alkali spin relaxation time is 1\,ms.}
\label{fig7}
\end{figure}

An example of the simulated Fourier spectrum by solving the above equations is shown in Fig.\,\ref{fig7}, where six additional resonance peaks appear between the He and Xe resonance peaks, in agreement with previous prediction according to Eq.\,\ref{fs}. Especially, to eliminate the residual phase information in Fig.\,\ref{fig7}(a), we plot Fig.\,\ref{fig7}(b) a Fourier spectrum mixed with realistic level of noise in the experiment \cite{Limes2018PRL}. When comparing the simulation spectrum with the experimental one, one shall remember that there might be slightly difference in the exact positions of the main two nuclear spin resonance peaks as the value of magnetic field $B_0$ in the experiment is not equal to 5\,mG as it is in the simulation. Although the parameters can not be exactly fitted to the experiment (for example, the He and Xe magnetization fields might be a bit higher than that expected in experiment), to a good confidence level, we can conclude that the multiple peaks spectrum agrees very well with the ``cross modulation'' spectrum observed in experiment \cite{Limes2018PRL}. 

\begin{figure}[H]
\centering
\includegraphics[width=\columnwidth]{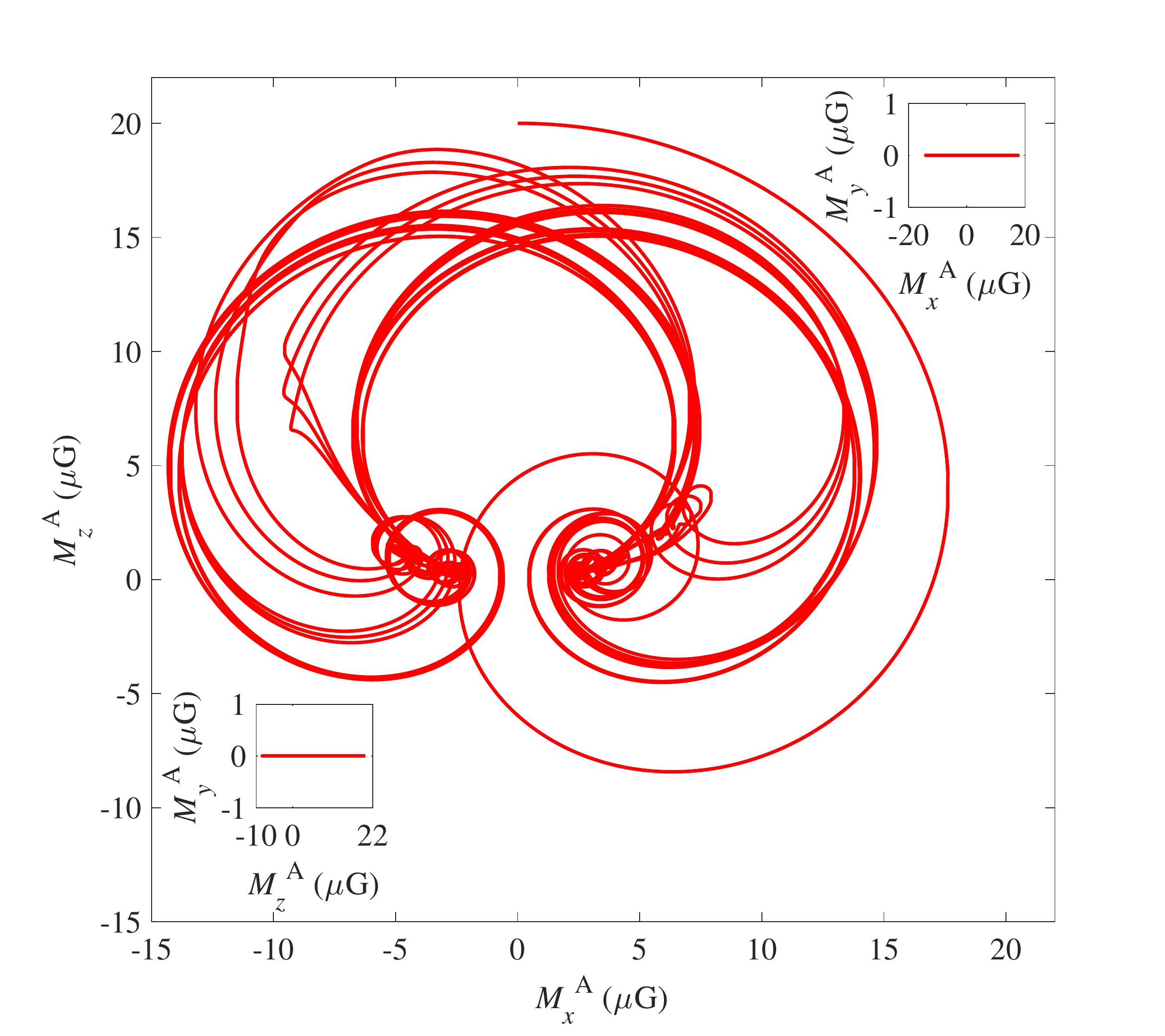}
\caption{Trajectory of alkali spin precession simulated for the restricted planar motion of partial coupling configuration. Two insets at bottom left and top right corners show a zero $M_y^{\rm A}$ component due to the restricted alkali spin precession within the $xz$ plane. The He and Xe initial magnetization fields and the alkali spin relaxation time are the same as in Fig.\,\ref{fig7}.}
\label{fig8}
\end{figure}

Interestingly, the multi-frequency spectrum of the tri-spin system on the condition of restricted planar motion resembles the multi-period solutions of the classical three-body dynamics for planar orbiting planets in astrophysics \cite{3body2013PRL}. In this sense, we can view the multiple frequency spectral components as multiple orbits of spin precession. For example, in Fig.\,\ref{fig8}, we plot the trajectory of alkali spin precession for the initial two seconds in a restricted $xz$ plane, according to Eq.\,\ref{cbe2}. One can find clearly spaced clusters of spin precession orbits. These multiple orbits correspond to spin precession with different periods or frequencies than the two main resonance frequencies, as already shown in Fig.\,\ref{fig7}. We may also notice that the precession orbits in Fig.\,\ref{fig8} show more patterns or asymmetry than that in Fig.\,\ref{fig4} and it is in consistent with a richer spectral composition while comparing the spectra.

The multi-frequency spectrum and multiple orbits of spin precession represent a new kind of instability, which is in fact a multistability with infinite number of spin precession orbits due to the everlasting changing mode of motions in three-body dynamics. As long as we work with such an $A$-$N_1$-$N_2$ tri-spin system in the partial coupling configuration (in order to remove the noble gas NMR shift due to alkali metal polarization), we can not remove the additional peaks in the multi-frequency spectrum from the root. However, we would like to point out that the presence of the instability does not necessarily mean the tri-spin system is no more suitable for precision measurement. Actually, to the best accuracy of the numerical simulation, there is no evidence that the main two nuclear resonance peaks shift due to the presence of the multiple resonance peaks.

\section{discussion and conclusion}

The agreement between our numerical simulation and experimental results leads us to the discovery of a new source of dynamical instability in the alkali metal-noble gas comagnetometry or gyroscope. Different from the $A$-$N$ dual-spin comagnetometry or gyroscope working at a compensated zero magnetic field, where the nonlinearity or dynamical instability happens due to the perturbation of an external large angle excitation pulse \cite{Kornack2002PRL}, here in the $A$-$N_1$-$N_2$ tri-spin system, the nonlinearity lies in the internal instability of three-body dynamics. 

The methods dealing with the instability are different. For the dual-spin gyroscope, one shall avoid or shield perturbation by external intense pulse of electromagnetic field in order to not driving the dual-spin system into its dynamically instable regime. For the tri-spin gyroscope, the only concern is that one shall make a compromise between large enough gyroscope signal and small enough tilting angle $\theta$ according to Eq.\,\ref{theta}. As long as we can distinguish the main two nuclear spin resonance peaks out of the spectrum with crowded multiple peaks, the tri-spin ensemble serves as a good system for precision rotation measurement according to Eq.\,\ref{rotation}. In this sense, the numerical simulation answers the concern quantitatively. 

In conclusion, we have studied the spin dynamics at different coupling configurations for the alkali metal-noble gas tri-spin NMR gyroscope. We report a new source of instability for the tri-spin NMR gyroscope, originating from the instability of three-body dynamics. The numerical simulation results agree well with recently reported experimental spectrum and enable a quantitative understanding of its complex structure. These findings can be helpful for exploiting the full potential of the tri-spin NMR gyroscope. 

The author would like to thank Dong Sheng and Mark Limes for helpful discussions. The author would also like to thank Andrei Ben-Amar Baranga for kind suggestions during the revision of the manuscript.

\section{appendix}
\begin{figure}[H]
\centering
\includegraphics[width=\columnwidth]{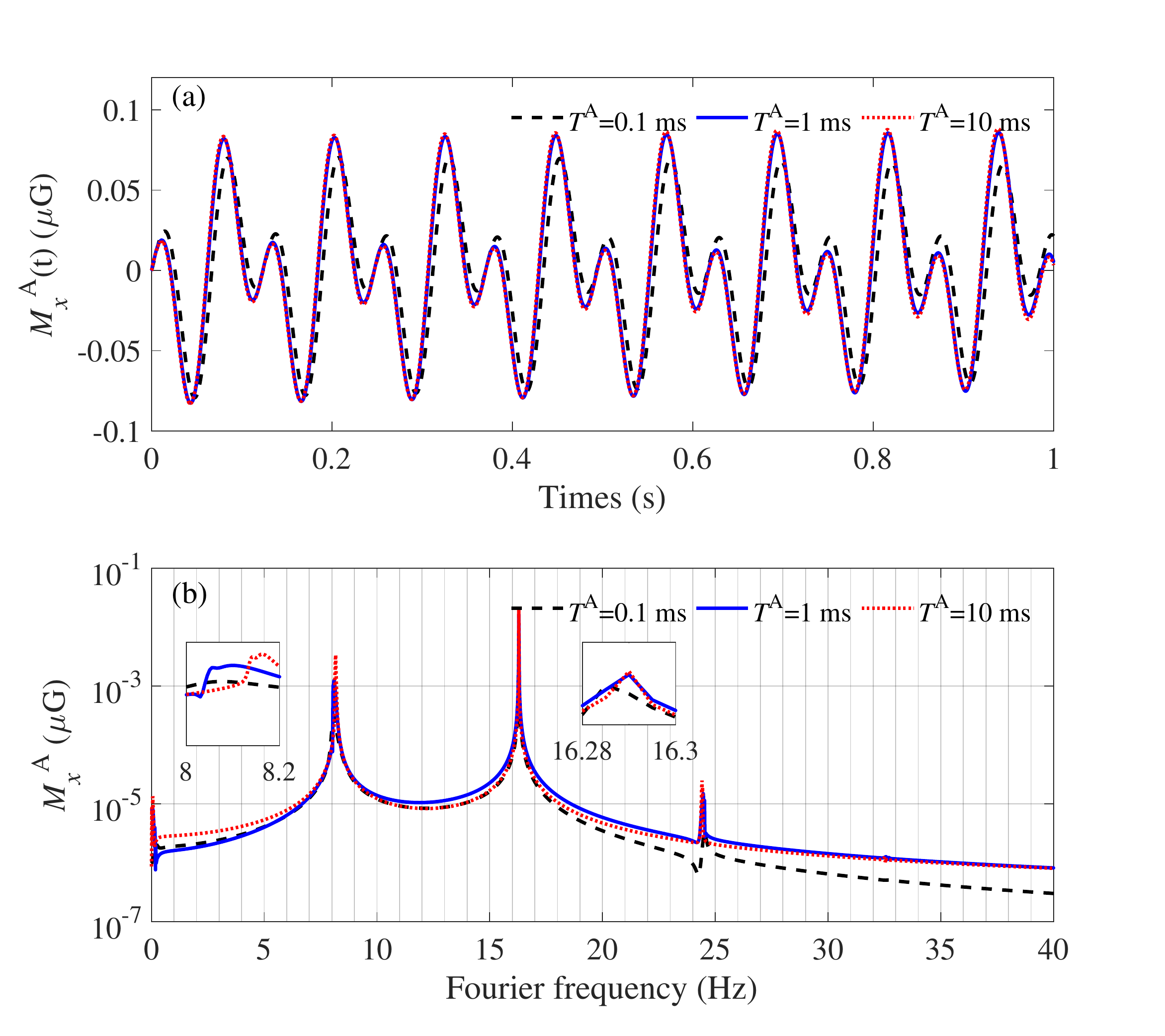}
\caption{Gyroscope signal versus time (a) and frequency (b) for different alkali spin relaxation times simulated for the complete coupling tri-spin system, with nuclear magnetization $\lambda M_0^{\rm N}$=0.1\,mG.}
\label{fig9}
\end{figure}

\begin{figure}[H]
\centering
\includegraphics[width=\columnwidth]{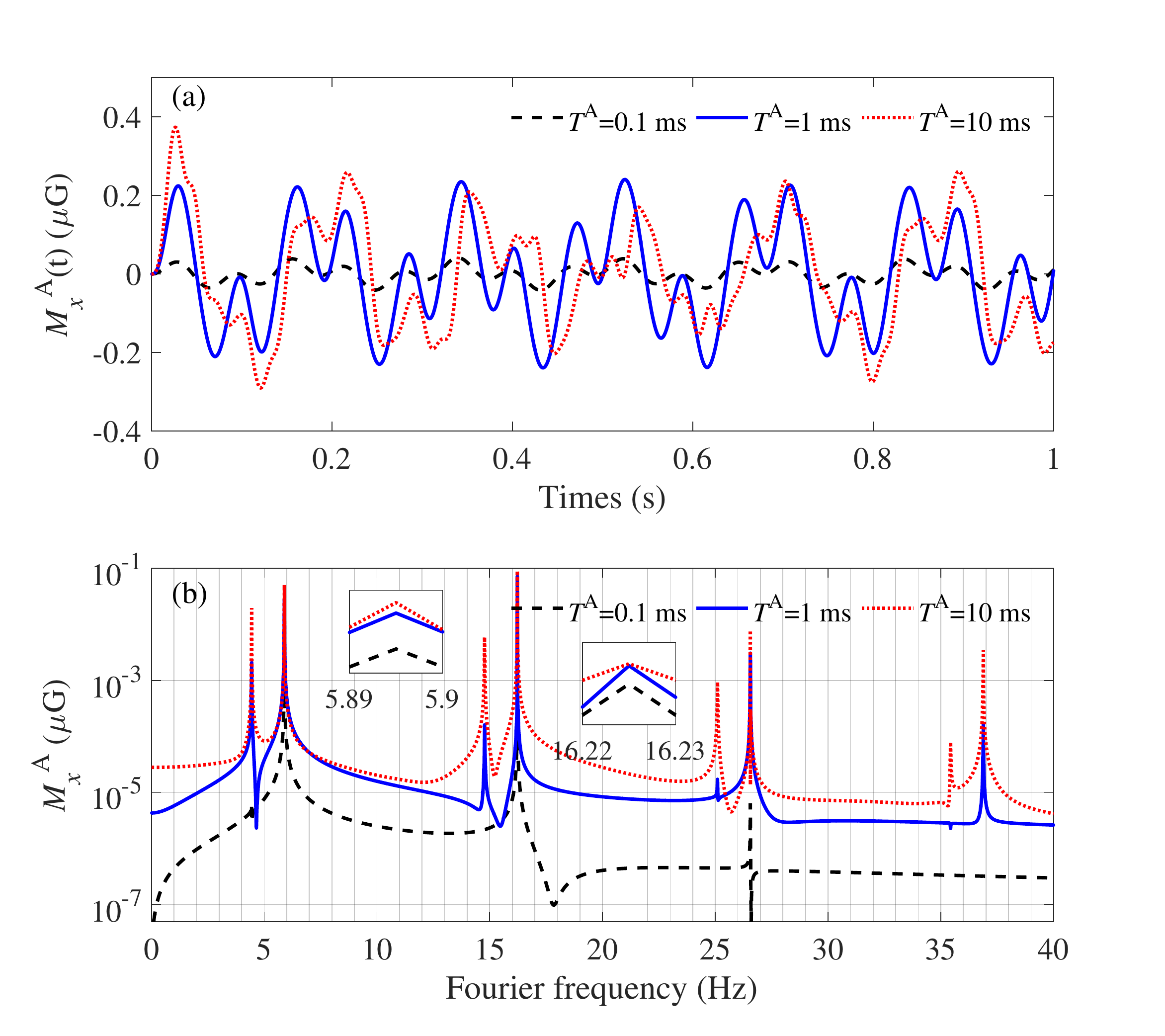}
\caption{Gyroscope signal versus time (a) and frequency (b) for different alkali spin relaxation times simulated for the partial coupling tri-spin system, with nuclear magnetization $\lambda M_0^{\rm N}$=0.1\,mG.}
\label{fig10}
\end{figure}

We also investigate the gyroscope response at different alkali spin relaxation times. Fig.\,\ref{fig9} and \ref{fig10} show gyroscope signal simulated for the complete and the partial coupling tri-spin systems, respectively.
 
\begin{figure}[H]
\centering
\includegraphics[width=\columnwidth]{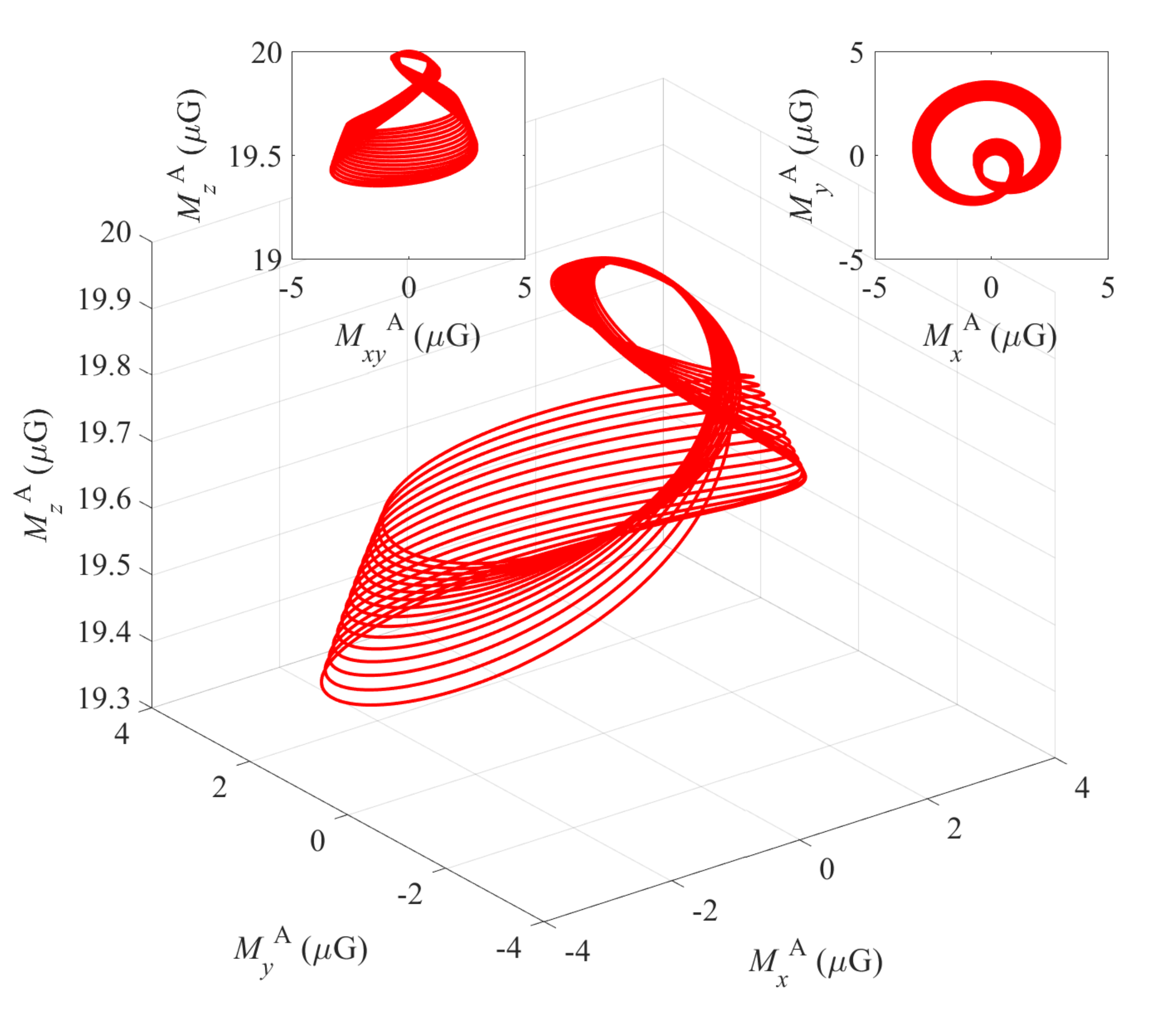}
\caption{Trajectory of alkali spin precession simulated for the complete coupling tri-spin system, with nuclear magnetization $\lambda M_0^{\rm N}$=0.5\,mG and alkali spin relaxation time $T^{\rm A}$=1\,ms.}
\label{fig11}
\end{figure}

\begin{figure}[H]
\centering
\includegraphics[width=\columnwidth]{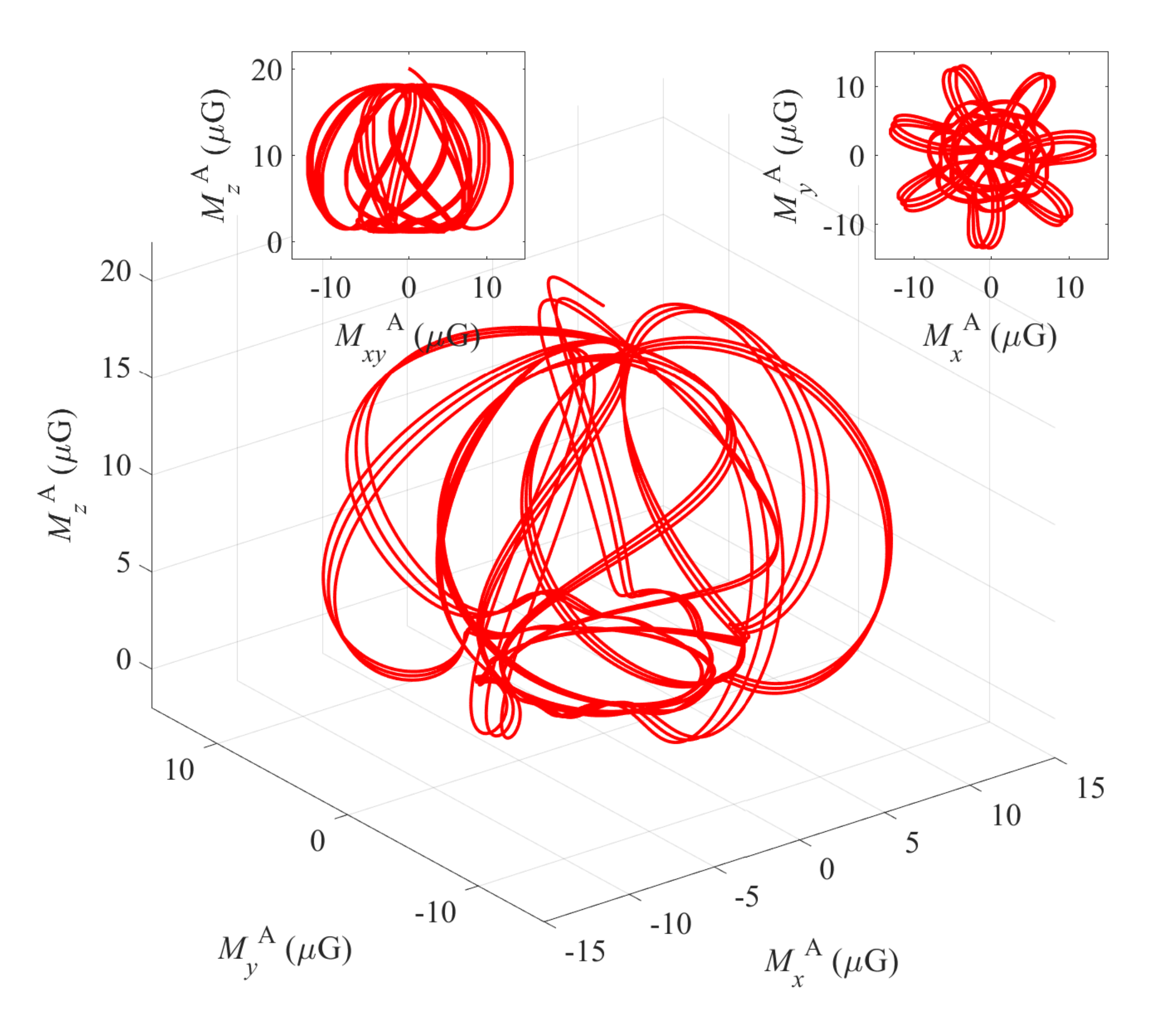}
\caption{Trajectory of alkali spin precession simulated for the partial coupling tri-spin system, with nuclear magnetization $\lambda M_0^{\rm N}$=0.5\,mG and alkali spin relaxation time $T^{\rm A}$=1\,ms.}
\label{fig12}
\end{figure}

We also simulated the trajectories of alkali spin precession at stronger nuclear magnetization. Fig.\,\ref{fig11} and \ref{fig12} show the trajectories for the complete and the partial coupling tri-spin systems, respectively. 

\end{document}